# Cures, Treatments and Vaccines for Covid-19: International differences in interest on Twitter

Mike Thelwall, Statistical Cybermetrics Research Group, University of Wolverhampton, UK. Email: m.thelwall@wlv.ac.uk Orcid: 0000-0001-6065-205X

Since the Covid-19 pandemic is a global threat to health that few can fully escape, it has given a unique opportunity to study international reactions to a common problem. Such reactions can be partly obtained from public posts to Twitter, allowing investigations of changes in interest over time. This study analysed English-language Covid-19 tweets mentioning cures, treatments, or vaccines from 1 January 2020 to 8 April 2021, seeking trends and international differences. The results have methodological limitations but show a tendency for countries with a lower human development index score to tweet more about cures, although they were a minor topic for all countries. Vaccines were discussed about as much as treatments until July 2020, when they generated more interest because of developments in Russia. The November 2020 Pfizer-BioNTech preliminary Phase 3 trials results generated an immediate and sustained sharp increase, however, followed by a continuing roughly linear increase in interest for vaccines until at least April 2021. Against this background, national deviations from the average were triggered by country-specific news about cures, treatments or vaccines. Nevertheless, interest in vaccines in all countries increased in parallel to some extent, despite substantial international differences in national regulatory approval and availability. The results also highlight that unsubstantiated claims about alternative medicine remedies gained traction in several countries, apparently posing a threat to public health.

## Introduction

At the start of the Covid-19 pandemic, it was unclear whether it would be amenable to any kind of medical intervention. It was not known whether it would be possible to develop a cure for it, in the sense of a medicine that would kill the virus. Similarly, it was unclear whether any Covid-19 vaccines could be developed. At the same time treatments, in the sense of medical interventions that would limit the damage caused by the virus (e.g., medicines reducing its virulence; ventilators keeping patients alive whilst their antibodies developed) became important to help people with symptoms from the virus. This article investigates whether it is possible to gain insights into international differences in attention to these three issues on Twitter. Explorations of international differences may shed light onto national beliefs that may influence the acceptance of health recommendations, such as vaccination, as well as adherence to safety measures, such as social distancing. This comparison may also give a new perspective from which to investigate a social aspect of the Covid-19 pandemic. Most people in most countries probably believe that inequalities in access to healthcare is unfair (von dem Knesebeck et al., 2016), despite substantial differences between and within countries (e.g., Balarajan, et al., 2011), giving an additional incentive for international comparisons.

The issue of vaccine hesitancy and the anti-vaccination movement in the West have highlighted that medical experts are not always trusted, whether due to conspiracy theories (Hornsey et al., 2018; Romer & Jamieson, 2020), concerns due to historical abuse by medical professionals (Green et al., 1997; Weindling, 2008) or other factors. A second issue is that traditional remedies may compete for allegiance with modern medicine (i.e., biomedicine), for people that cannot afford, or live in a remote rural area without access to, biomedicine (Sen & Chakraborty, 2017). Alternative therapies, such as homeopathic remedies, are also sometimes accepted as a worldview choice, out of fear of the side-effects of pharmaceuticals, for specific ailments (Astin, 1998; Beer et al., 2016), for cultural reasons (Lee, et al., 2000; Rudra, et al., 2017), because it is effective in some contexts (WHO, 2013), or following bad experiences with conventional doctors (Avina & Schneiderman, 1978). In the USA, more educated (and presumably richer) people are more likely to seek alternative medicine (Astin, 1998). Individuals may also accept some medical developments and reject others for ethical or personal reasons. For example, the use of genetics to estimate disease susceptibility is not universally accepted (Zhang et al., 2021). Citizens may also try both alternative and conventional medicines in the hope that one of them works and neither are harmful, and doctors may sometimes recommend traditional therapies that were previously regarded as not evidence-based, such as acupuncture or yoga, including herbal medicines in India (Sen & Chakraborty, 2017).

Alternative medicines can be widely accepted or have an official status in some countries. Their governments may oversee them to some extent, as in Saudi Arabia (Aboushanab & Baslom, 2021), and may also encourage them within limits. For example, they are government-approved in India, including Ayurveda, Yoga and Naturopathy, Unani, Siddha, Sowa-Rigpa and Homeopathy (AYUSH) (Rudra, et al., 2017). These are overseen through the Ministry of Ayush (www.ayush.gov.in). On 17 April 2021, a ministry information leaflet stated that Ayush treatments could aid symptom management but did not cure Covid-19 (www.ayush.gov.in/docs/faq-covid-protocol-Revised.pdf). Nevertheless, an alternative medicine company in India is marketing a herbal remedy for Covid-19 called Coronil. Its efficacy is supported by one small (95 patients) randomised placebo-controlled trial (https://doi.org/10.1016/j.phymed.2021.153494), but it seems likely to be ineffective.

Given the known international differences in attitudes towards medicine, the dual aims of this article are to assess whether tweets can shed light on international differences in attitudes towards potential biomedical or alternative Covid-19 cures, treatments, and vaccines, and to seek insights into international differences in attention given to these three issues during the pandemic. Social media are known to be a practical source of rapid public opinion information (Bengston et al., 2009; Karami, et al., 2018; Tavoschi, et al., 2020), although it is imperfect because not everyone uses all sites and not all topics are posted about. The focus is on Twitter for pragmatic reasons: it is a public source of news, politics, academic and other information that is extensively used in many countries and is available to be mined for academic research. Although other social media sites have more users (e.g., Facebook, YouTube, Weibo), none have these properties. The focus is on English to allow direct international comparisons and because English is a language of news due to the USA's importance for the media marketplace (e.g.., Chang, 1998), including on Twitter (Wilkinson & Thelwall, 2012). The countries to be compared is a pragmatic choice: those posting the most tweets in English about Covid-19. This method has many limitations that are analysed as part of this study for the first objective.

## Methods

The research design was to gather a large sample of Covid-19 tweets in English and compare international daily trends in volume of tweets over time, reading the tweets to find reasons for countries deviating from the norm. The tweets were gathered in two stages. First, Twitter was monitored at the maximum rate permitted by the free Applications Programming Interface from 10 March 2020 to 8 April 2021 using the following queries, specifying English as the language: covid-19, covid19, coronavirus, and "corona virus". Second, the same queries were run for 1 January to 9 March 2020 through the academic access to the historical Twitter archive, as released in January 2021. This allowed the current study to cover the entire period of the pandemic.

Although Twitter removes some spam from its feed and archive, the tweets gathered were pre-processed to reduce the influence of any remaining spam and prevent the results from being dominated by individual prolific tweeters. For this, the tweets were first processed to remove duplicates and near duplicates (tweets that were identical except for @usernames). A common spamming trick is to send the same message multiple times, targeting different users, and this remove most such messages. The second stage was to allow a maximum of one tweet per user per month. This reduces the influence of prolific tweeters, with each user allowed a maximum of 16 tweets (one per month). This extra stage allows the results to reflect average contributions of users rather than the energy of prolific individuals. This is also an anti-spam step since bots can be prolific, for example by automatically tweeting a range of news sources. After this stage, there were 26,266,178 non-duplicate English Covid-19 tweets.

Twitter allows users to report their location, with many users declaring a country, city, or town. These country declarations and the names of large distinctive cities (e.g., New York, Lagos, Frankfurt) were used to assign users to nations. Users not declaring a country or distinctive city name (49.9%) were ignored. The 32 countries with the most tweets were selected for analysis. It is not useful to include all countries because those with few tweets would not show daily trends. The cut-off at 32 was chosen to include the 32$^{nd}$ country, Sweden, which took a different approach to the pandemic social distancing to the rest of Europe. Sweden did not return any interesting trends in this study, however, so the set analysed could safely have been smaller.

A graph was plotted of the median of the national percentages of tweet about cures, treatments and vaccines to report the overall average international level of interest. Using the median of the countries rather than the overall average for all tweets prevents the results from being dominated by the country with the most tweets, the USA.

Graphs were plotted of the daily percentage of cure (cure, cured, curing), treatment (treat, treating, treated, treatment) and vaccine (vaccine, vaccinated, vaccinating) tweets for each country, based on the percentage of matching tweets. This is a highly simplistic approach to track interest in these three issues because these terms can be used in other contexts and other terms can be used to describe the issues. In particular, vaccines, treatments and cures might be referred to by name rather than with the generic term, and the extent to which this occurs is likely to vary substantially over time and between countries. For example, more recent vaccine-related tweets are more likely than early tweets to mention individual vaccine names.

Human Development Index scores were also compared against the results to give a simple impression of whether levels of interest could be related to the level of human development in each country. Although there are many national factors that could be compared, such as levels of interest in alternative medicine, human development scores have a relatively robust source and are regularly calculated by the United Nations.

## Results

The raw data behind all graphs is in the online supplement (https://doi.org/10.6084/m9.figshare.14446923). There were similar low median levels of interest in treatments, cures, and vaccines in English Twitter, across the 32 countries investigated until July (Figure 1). The Russian Sputnik V vaccine regulatory approval in Russia triggered a large spike and was part of a moderate increase in interest in vaccines. This moderate increase in interest before and after the Sputnik V spike was mainly due to news of vaccine trial progress for the Oxford AstraZeneca (UK), Moderna (USA), Covaxin (India) and Pfizer-BioNTech (Germany) vaccines. The Pfizer-BioNTech vaccine phase 3 trial success subsequently created by far the largest spike, leading to a substantially higher level of interest in vaccines. This increased steadily afterwards, with spikes presumably associating with similar announcement for other vaccines as well as regulatory approval and rollout announcements. This example illustrates that social media data is particularly useful for identifying trends in public interest retrospectively (Thelwall, 2007), confirming that there was a low level of interest in vaccines in the early stages of the pandemic.

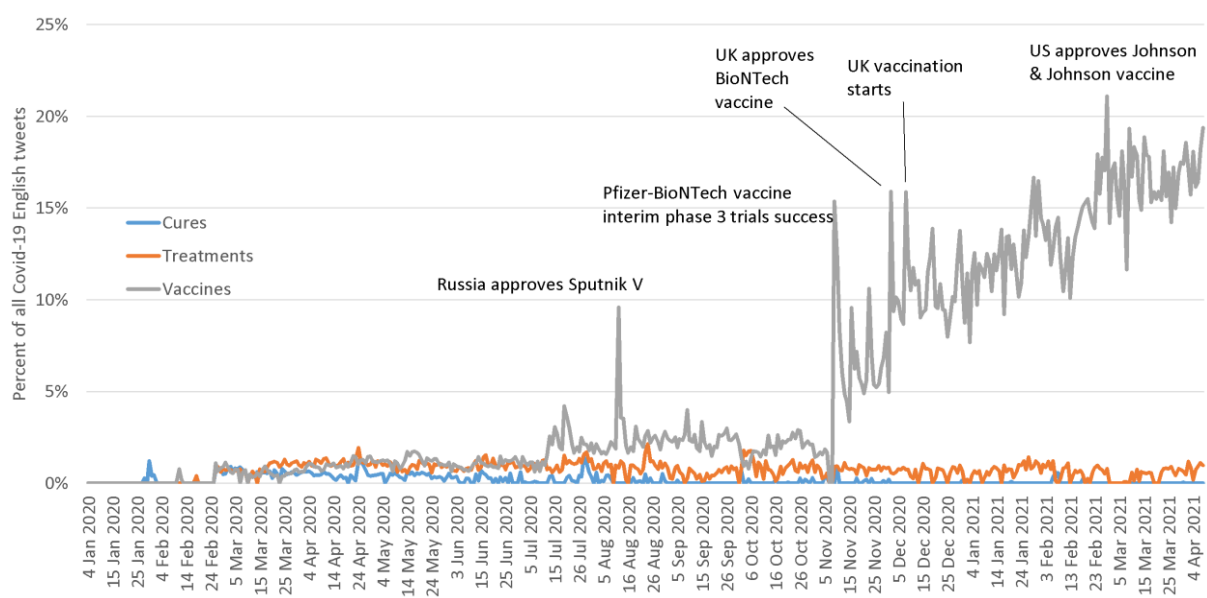

Figure 1. Median national percentages of English Covid-19 tweets mentioning cures, treatments, or vaccines. The largest spikes are labelled with the main topic on the day.

### National differences

National differences are compared through smoothed graphs (7 day moving averages) rather than exact graphs (as in Figure 1) because many of the countries had very jagged trend lines due to small sample sizes. Seven day averages were used to avoid weekend effects compared to, for example, five day averages. From the international comparison perspective, interest in Covid-19 cures decreased over time, with a generally low level of interest after April 2020 (Figure 2). Spikes in the graphs tend to indicate either variability due to low numbers of tweets or news stories associated with cures, usually of a local nature.

Nigeria seemed to have a particular interest in chloroquine and hydroxychloroquine, possibly influenced by Donald Trump's claims about them for the first spike. The second chloroquine spike from Nigeria was directly influenced by Cameroonian-American Dr Stella Immanuel's emphatic video claim at a right-wing event in the USA that hydroxychloroquine

(HCQ), zinc and azithromycin worked because none of her 350 patients had died. Her video was subsequently removed by Twitter as misinformation. It is not clear why Nigeria was apparently more interested in this than any other country in the sample. One of the other two spikes with more than 10 tweets was for a herbal remedy from Madagascar being promised to Nigeria. The remaining 10+ tweet spike was for the launch in India of the herbal Ayurvedic medicine Coronil as a remedy for Covid-19. This spike is perhaps surprising, considering that only 7% of Indians seeking outpatient care try Ayush services instead of biomedicine (Rudra, et al., 2017). Thus, only three claimed cures gained substantial attention in the data set, but it is not clear why Nigeria was most interested in cures, or whether any of the spikes were caused by sarcasm rather than serious interest in the topic.

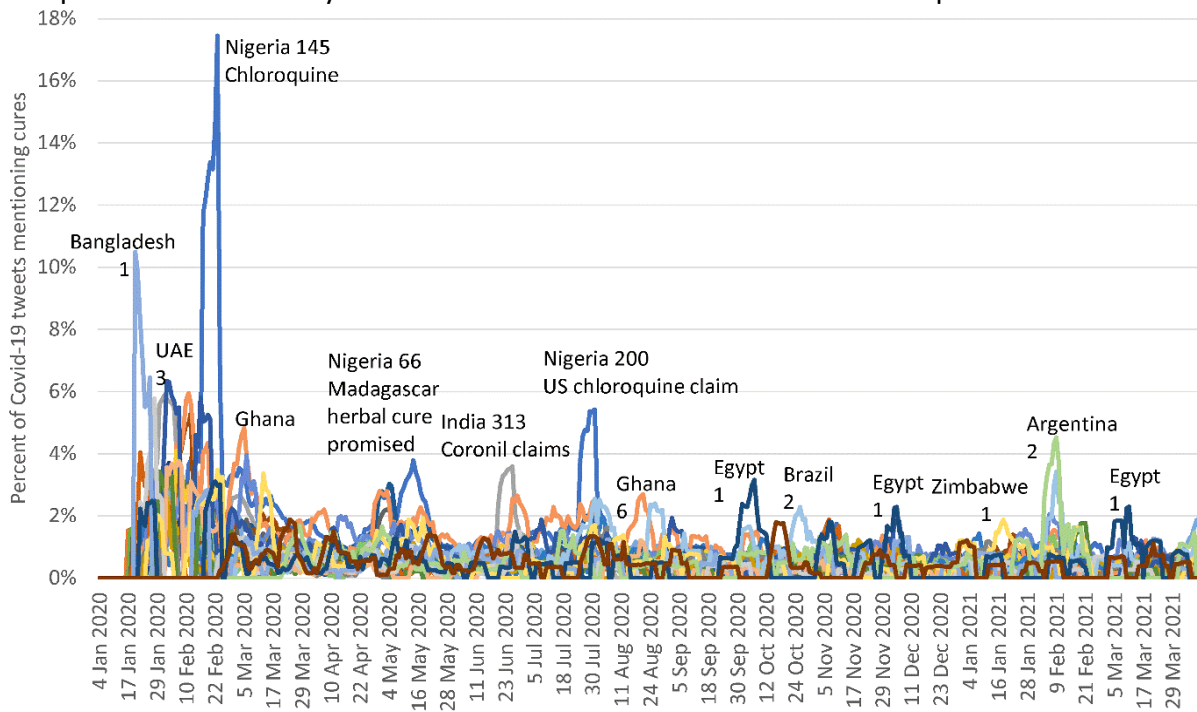

Figure 2. Covid-19 English Tweets containing cure(s), cured, or curing from 1 January 2020 to 8 April 2021 by country. Lines indicate 7-day moving averages. Numbers indicate tweets on the highest volume day and descriptions indicate topics when there are at least 10 tweets. Labelled graphs for each country are in the online supplement (https://doi.org/10.6084/m9.figshare.14446923).

Interest in Covid-19 treatments was approximately constant over time from April 2020 onwards, but tended to decrease in the second half (Figure 3, see also Figure 1). Large spikes in the graphs tend to indicate either variation due to small sample sizes or local news stories associated with treatments. The main exception was early interest in Pakistan about Chinese health workers travelling to Wuhan to help treat patients, risking death in the pandemic. All other spikes reflect local news stories. These were national treatments (United Arab Emirates, India), a national treatment recommendation (Brazil) and a domestically developed therapeutic drug starting trials (Uganda).

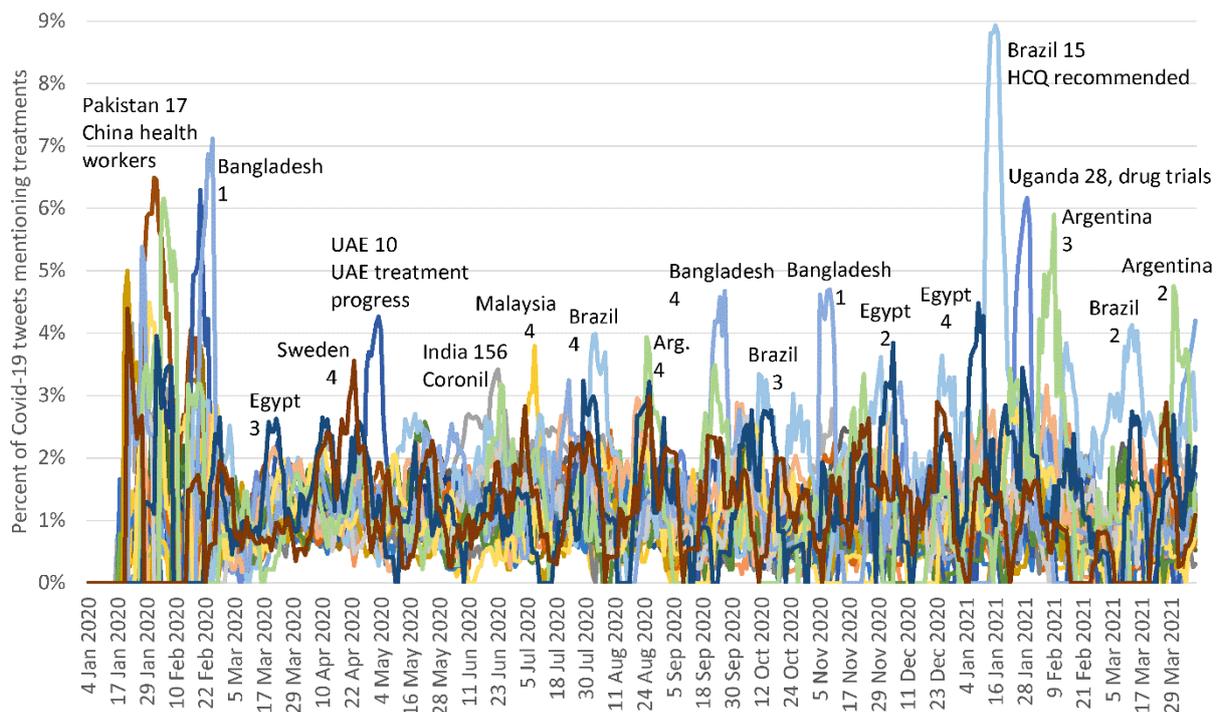

Figure 3. Covid-19 English Tweets containing treat(s), treatment(s), treated, or treating from 1 January 2020 to 8 April 2021, by country. Numbers indicate tweets on the highest volume day and descriptions indicate topics when there are at least 10 tweets. Lines are smoothed with a 7-day moving average. Labelled graphs for each country are in the online supplement (https://doi.org/10.6084/m9.figshare.14446923).

There was little interest in Covid-19 vaccines until November 2020, then constantly increasing in all countries (Figure 4). Large spikes in the graphs tend to indicate variation due to little data (<10 tweets) or national news. The graph did not exhibit synchronised spikes of interest on the dates when the first few vaccines were revealed to be effective through double-blind randomised clinical trials or the Sputnik V spike found in Figure 1. These spikes in the exact data of Figure 1 are smoothed in the 7-day moving average of Figure 4, which was necessary due to the small daily sample sizes for some countries. Nevertheless, these announcements seemed to trigger an overall relatively steady increase in interest in vaccines, with national spikes mainly associating with national vaccination drives. The two international stories generating spikes were news of a trial death in Brazil (biggest spike in India) and the main Pfizer-BioNTech vaccine announcement, triggering the biggest spike in Argentina.

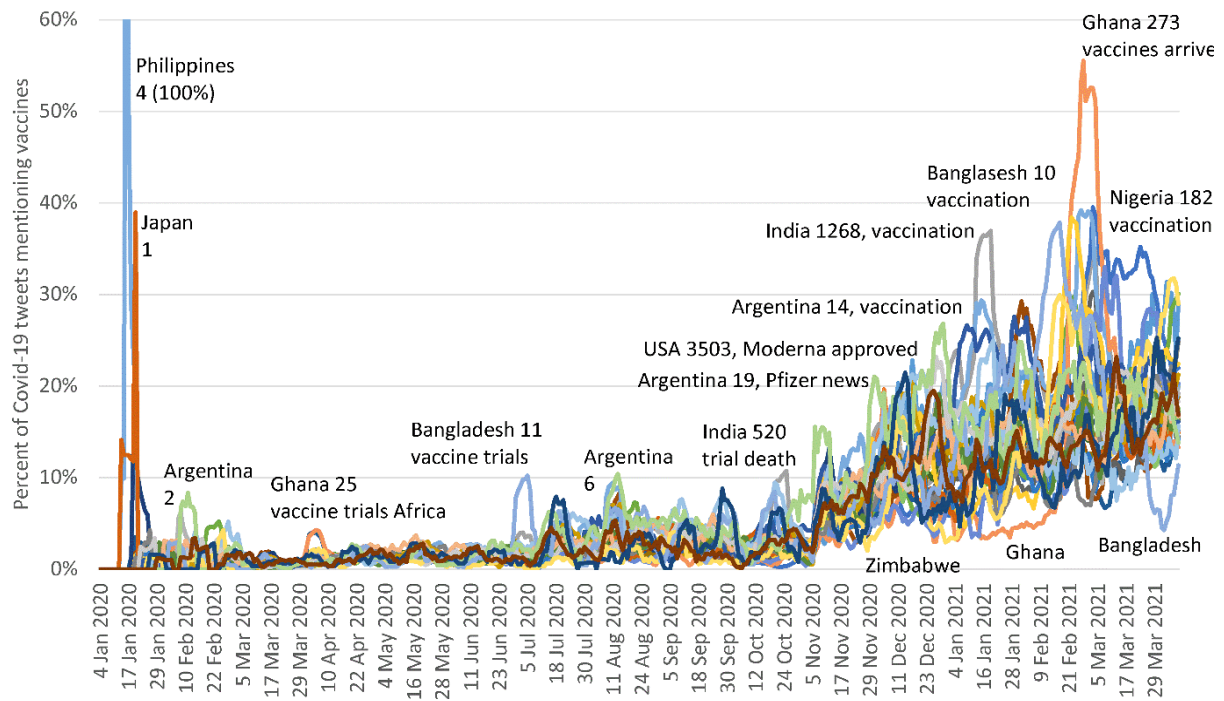

Figure 4. Covid-19 English Tweets containing vaccine(s), vaccinated, vaccination, vaccinating from 1 January 2020 to 8 April 2021, by country. Numbers indicate tweets on the highest volume day and descriptions indicate topics when there are at least 10 tweets. Lines are smoothed with a 7-day moving average. Labelled graphs for each country are in the online supplement (https://doi.org/10.6084/m9.figshare.14446923).

## *Interest level vs. human development*

The relative attention to cures, treatments and vaccines in each country was correlated against the human development index (HDI) 2020 score of each country (UN, 2020) to assess whether development level might be an influence. Such a relationship is a logical possibility because access to health care is a component of high human development in a society.

There was a strong negative relationship between HDI and percentage of cure tweets (Pearson correlation: -0.713, n=32): Cure tweets were more prevalent in countries with low HDI scores (Figure 5). This may be because poorer and lower HDI countries tended to have more flourishing herbal and other traditional remedies industries. With lower standards of scientific proof in this sector than for modern biomedicines, it is easier to claim to have created a herbal remedy than a biomedical cure, or there is a larger audience for claim-based products. For example, the president of Madagascar (a country not in the sample, and with a low HDI) was able to claim that a herbal drink, Covid-Organics, was a vaccine and cure for Covid-19, mandating it for Madagascan schoolchildren and exporting supplies to some African countries (AfricaNews, 2021). From a biomedicine perspective, a Covid-19 cure is a possibility because other viruses have cures, including the herb-based quinine cure for Malaria (a component of the Madagascar remedy). The method by which quinine works is not fully understood, so it is plausible that quinine or another herb extract could cure Covid-19, for example by preventing it from replicating (see also: Vellingiri, et al., 2020). Perhaps the highest profile herbal remedy for Covid-19 is Coronil from India (IndiaToday, 2020). Nevertheless, at the time of writing, it seems likely that neither worked, and insufficient evidence has been presented for either to be accepted as a treatment, vaccine or cure.

It is perhaps surprising that India's neighbour Bangladesh is an outlier for its relative absence of interest in cures, and another neighbour, Pakistan, also shows relatively little interest for its HDI. Given religious backing for the Indian remedy, it is possible that strong alternative medicine claims are less compatible with Islam than with Hinduism due to the roots of modern science being in Islamic nations (Iqbal, 2007). Nevertheless, alternative medicines are well known and used in both Bangladesh (Saha, et al., 2017) and Pakistan (by 52% in one study: Shaikh, et al., 2009), so religion-based explanations are too simplistic, and it is not clear why other countries have not discussed high profile Covid-19 remedies, at least on Twitter.

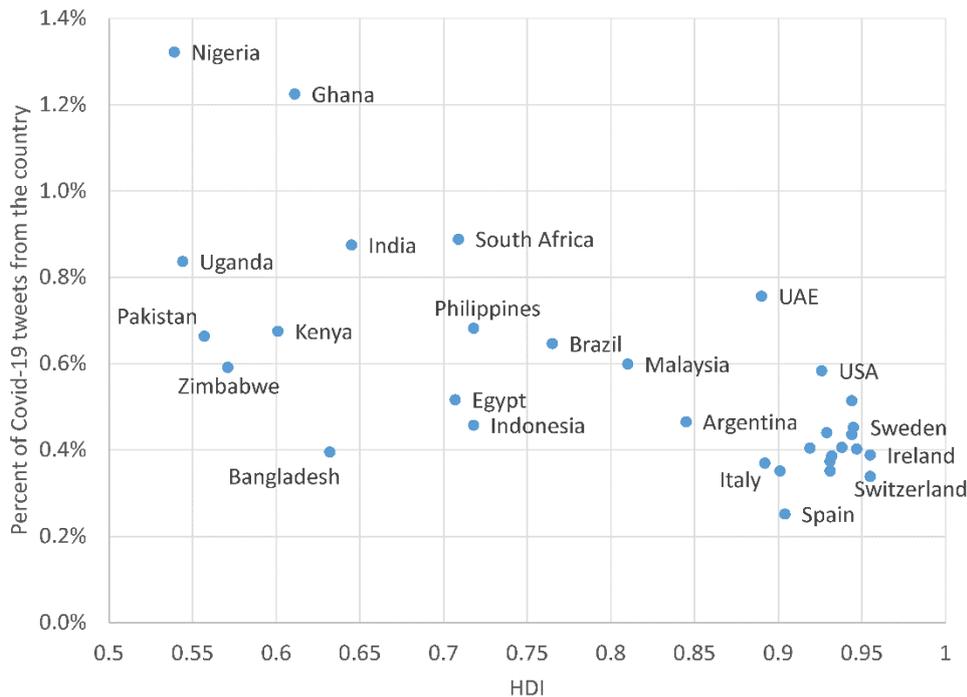

Figure 5. Percentage of cure tweets within the Covid-19 set against HDI for 32 countries (Pearson correlation: -0.713, n=32).

There was little relationship between HDI and level of interest in treatments (Figure 6). Tweets about treatments might be more expected in countries with high Covid-19 case rates because these have more people to be treated. This high incidence set excludes most of Africa, except South Africa, so the absence of a positive relationship in Figure 5 is surprising. Some of the outliers are consistent with the prevalence of Covid-19 cases. For example, Brazil and India (top of Figure 6) had high rates in by April 2021 and Ghana, Indonesia and New Zealand had low rates (JHU, 2021). There are also exceptions, however, with Italy (early on) and South Africa having many cases but few explicit mentions of treatment. The term treatment was often used to discuss suggested medicines, such as chloroquine (e.g., in Brazil) rather than ventilation or other treatments, however, skewing the results.

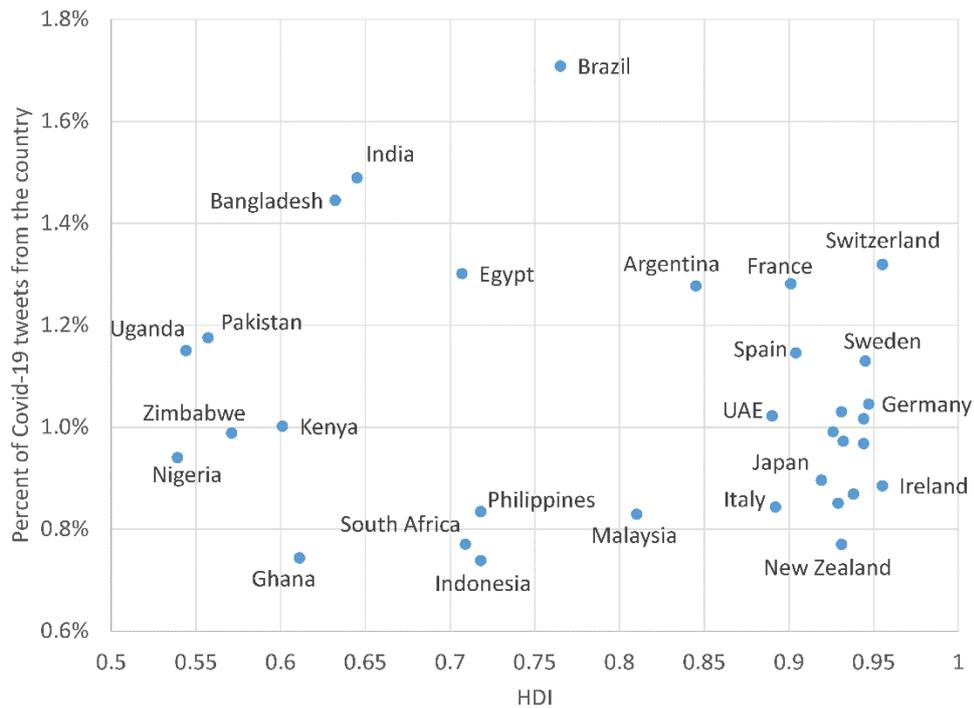

Figure 6. Percentage of treatment tweets within the Covid-19 set against HDI for 32 countries (Pearson correlation: -0.141)

There was a very weak tendency for higher HDI nations to tweet more about vaccines (Figure 7). More interest might be expected in countries that make it (e.g., India, USA, UK, Belgium) due to press coverage of national attempts to produce a vaccine. Similarly, more interest might be expected in the richer and higher HDI countries that vaccinated first and in large numbers (e.g., USA, UK, UAE) due to coverage of the vaccination programme.

The Figure 7 outliers are surprising. During the data collection period, the Philippines had vaccinated a lower proportion of its population than the world average (it started on 1 March 2021), although it was unusual in authorising products from four different countries (UK, USA, China, Russia) so this variety may have generated extra interest. From relevant Philippine tweets, there did not seem to be any other Philippine-specific issue. It is difficult to find out why there was relatively little vaccine-related tweeting in South Africa because the absence of tweets does not give a clue to the reason. Moreover, South Africa had additional vaccine news compared to most other countries due to discussions about whether the AstraZeneca vaccine worked for the South African Covid-19 variant and the country's offer to swap it for another vaccine. From the time series graph, interest in vaccination in South Africa declined after this event, so the low numbers of tweets may reflect delayed mass vaccination.

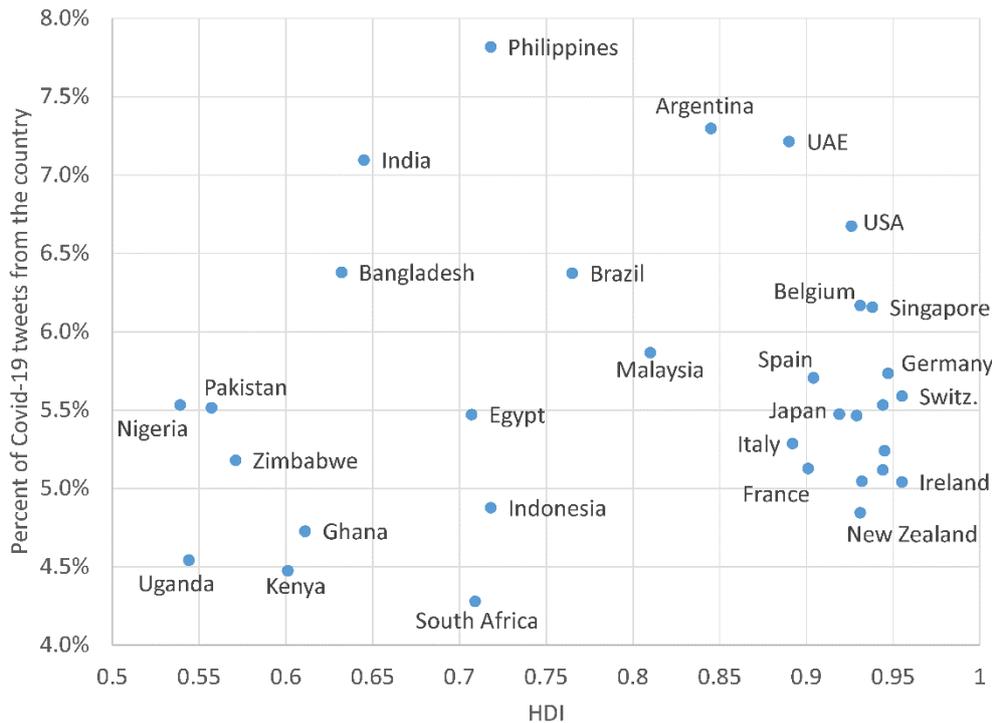

Figure 7. Percentage of vaccine tweets within the Covid-19 set against HDI for 32 countries (Pearson correlation: 0.112)

## Limitations

The results have several substantial limitations and so should be interpreted as suggestive of trends rather than strong evidence of them. An important limitation is that the focus the small sets of keywords used do not capture all mentions of cures, treatments, or vaccines, given that they may be mentioned by name (e.g., Coronil), through phrases (e.g., the new medicine should help the most critically ill), or through anaphora (I got it injected in my arm today!) A second important limitation is that Twitter users are not representative of the general population in any country, and the demographic of users may vary by country. This is exacerbated by the focus on English, which is not the main spoken language of most countries analysed. Related to this, tweeters may focus on the news rather than their own concerns and may employ sarcasm or irony to discuss the pandemic, giving a false impression of serious interest. The low numbers of daily tweets for some countries are also a methodological problem, delivering apparently false spikes in some graphs. The smoothing in the national comparison charts that was used to reduce their spikiness also tended to obscure the genuine spikes by spreading them over a week instead of a day.

## Conclusions

From a methodological perspective, the results show that Twitter can give plausible suggestions about trends and international differences related to the current pandemic, although the limitations discussed above mean that they must be interpreted cautiously. At the moment, Twitter seems to be the only practical free source of sufficient large-scale news-related content from the public to run this type of analysis. In theory, it would be possible to run similar larger scale analyses with data purchased from other international source, such as Facebook, Instagram and Weibo, however.

In terms of the overall trends (Figure 1), the results confirm that cures have never been a major topic of discussion for Covid-19. Perhaps surprisingly, treatments have also never been a main topic despite a variety of them being constantly in use throughout the pandemic and little apparent early hope for cures or vaccines. In the latter case, it seems likely that treatments were discussed by name (e.g., ventilators, paracetamol) rather than characterised explicitly as (partial) treatments. The sudden November spike for vaccines suggests that the possibility of a successful vaccine was a surprise, despite some prior interest from vaccine trials news and Sputnik V in Russia. The substantial and continued increase in interest in vaccines confirms that they seem to be widely internationally accepted as the key Covid-19 development at the time of writing (April 2021). In particular, they have not been ignored even in countries where there are claimed alternative cures have been sold. From a current public health perspective, it is good news that none of the countries analysed are ignoring vaccines.

From an international comparisons perspective (Figures 2-4), it is unsurprising that when countries deviate from the average with a spike of interest then this is usually due to a topic of national concern. These topics include progress news about locally developed cures or vaccines as well as more concrete events, such as the introduction of new products or vaccines into the country, or national regulatory approval. These events also suggest that early tweets have often been forward looking, reporting progress towards successful vaccination perhaps in the hope of an eventual escape from the pandemic in the absence of more tangible successes.

The main finding about the relationships between national human development and interest in cures, treatments and vaccines is that potential cures seemed to attract more interest in less developed countries, perhaps because of a greater tradition of relying on cheap herbal remedies by people unable to afford modern biomedicine. This perhaps underlines the greater threat posed by Covid-19 in contexts where modern evidence-based biomedicine is not the cultural norm, perhaps including deprived areas or poorer segments of society in richer nations. Whilst such remedies can play a valuable role in national healthcare systems (WHO, 2013), it seems clear that incorrect claims of cures, treatments or vaccines may have fatal consequences during pandemics. The "ability to control and regulate [modern and alternative medicine] advertising and claims" (WHO, 2013, p. 12) has already been recognised by the World Health Organisation but the Covid-19 situation highlights the need to ensure that strong claims are not made for traditional therapies without high standards of evidence. It also emphasises the importance of obtaining acceptance for vaccination programmes in the multiple different cultural contexts of the world to ensure that the disease is eradicated or controlled, assuming that current and future vaccines are effective.